\documentclass[aps,prd,onecolumn,superscriptaddress,nofootinbib,floatfix,11pt]{revtex4-2}

\usepackage{graphicx} 
\usepackage{xcolor}
\usepackage{subfigure}
\usepackage{multirow}
\usepackage{float}
\usepackage{amsmath}
\usepackage{amssymb}
\usepackage{bm}
\usepackage{slashed}
\usepackage{braket}
\usepackage{breqn}
\usepackage{verbatim}
\usepackage[utf8]{inputenc} 
\allowdisplaybreaks
\usepackage[
colorlinks=true,
linkcolor=blue,
breaklinks=true,
urlcolor=blue,
citecolor=blue]{hyperref}
\usepackage{appendix}
\usepackage{orcidlink}

\newcommand{\pku}{\affiliation{School of Physics, Peking University, Beijing 100871, China}}

\newcommand{\ucas}{\affiliation{School of Physical Sciences, University of Chinese Academy of Sciences, Beijing 100049, China}}

\newcommand{\thu}{\affiliation{Department of Physics and Center for High Energy Physics, Tsinghua University, Beijing 100084, China}}

\begin{document}
\title{A possible $\Sigma^*$ or $\Lambda^*$ resonance with $J^P=3/2^-$ in $K^-p\to K\Xi$ scattering}

\begin{abstract}
We analyze the $K^-p\to K^+\Xi^-$ and $K^-p\to K^0\Xi^0$ processes in the energy region $1.8<\sqrt{s}<2.8$ GeV within an effective Lagrangian approach. 
The $\Lambda(1800)$ and $\Sigma(2250)$ resonances, along with the ground states $\Sigma$ and $\Lambda$, are included.
Additionally, a possible $J^P=3/2^-$ $\Sigma^*$ or $\Lambda^*$ resonance with a mass around 1.9 GeV and a width of approximately 200 MeV is introduced to describe the structure at 2.0 GeV in the total cross section and reproducing the threshold behavior. 
The two possible solutions corresponding to $\Sigma^*(3/2^-)$ and $\Lambda^*(3/2^-)$ cannot be distinguished by the existing data. 
Predictions for the polarization of the final-state $\Xi$ and the cross section of $K^-n \to K^0\Xi^-$ are compared with the experimental data, we find that the results of solution-II with $\Lambda^*(3/2^-)$ are much better.
We also discuss the possible interpretations of the introduced $3/2^-$ hyperon as a pentaquark candidate, e.g. an $S$-wave $K\Xi(1530)$ hadronic molecule.
However, since the polarization data suffer from rather large uncertainties, more data inputs are needed in future experiments, for example, J-PARC, HIAF and JLab.
\end{abstract}

\author{Zheng-Li Luo\orcidlink{0009-0001-7326-634X}}\email{luozhengli@stu.pku.edu.cn}
\pku

\author{Jia-Jun Wu\orcidlink{0000-0003-4583-7691}}\email{wujiajun@ucas.ac.cn}
\ucas 

\author{Bing-Song Zou\orcidlink{0000-0002-3000-7540}}\email{zoubs@mail.tsinghua.edu.cn}
\thu


\maketitle

\section{Introduction}

Hadron spectroscopy is a crucial avenue for understanding the non-perturbative nature of the strong interaction. 
While the constituent quark model successfully describes the ground-state baryons~\cite{Gell-Mann:1964ewy, Zweig:1964ruk}, many excited states predicted in the mass region around 2 GeV remain poorly established experimentally, especially in the hyperon sector~\cite{ParticleDataGroup:2024cfk}. 
In addition to conventional three-quark configurations, alternative pictures such as hadronic molecules have also been proposed~\cite{Guo:2017jvc}. 
The $K^-p \to K\Xi$ scattering process provides a golden platform for exploring $\Sigma^*$ and $\Lambda^*$ resonances. 
Precise knowledge of these resonances and their couplings to the $K\Xi$ channel is essential for testing the predictions of the constituent quark model for the strangeness $S=-1$ hyperon spectrum~\cite{Capstick:1986ter}, probing possible exotic configurations beyond the conventional $qqs$ picture~\cite{Guo:2017jvc}. 

The available experimental data for the $K^-p \to K\Xi$ reaction in the energy region $1.8<\sqrt{s}<2.8$ GeV date back to bubble chamber experiments performed in the 1960s and 1970s. 
For the $K^+\Xi^-$ channel, total cross section data are reported in Refs.~\cite{Cooper:1962rxa,Carmony:1964zza,Badier:1965zzc,Berge:1966zz,Birmingham-Glasgow-LondonIC-Oxford-Rutherford:1966onr,London:1966zz,Trippe:1967wat,Burgun:1968ice,Trower:1968zz,Dauber:1969hg,Griselin:1975pa,Briefel:1977bp,deBellefon:1977sw}, while differential cross section data are given in Refs.~\cite{Birmingham-Glasgow-LondonIC-Oxford-Rutherford:1966onr, London:1966zz, Trippe:1967wat, Burgun:1968ice, Trower:1968zz, Dauber:1969hg}. 
For the $K^0\Xi^0$ channel, total cross section data are reported in Refs.~\cite{Badier:1965zzc, Berge:1966zz, Burgun:1968ice, Dauber:1969hg, Carlson:1973td, Briefel:1977bp, deBellefon:1977sw}, and differential cross section data in Refs.~\cite{Berge:1966zz, Burgun:1968ice, Dauber:1969hg, Carlson:1973td}. 
Notably, the $K^+\Xi^-$ channel contains considerably more data points than the $K^0\Xi^0$ channel. 
In addition, differential cross sections are available at only a few energy points, in contrast to the relatively more abundant total cross section data. 
The existing measurements suffer from limited statistics 
and sizable uncertainties, which impose significant limitations on precise theoretical analyses. 

The $K^-p\to K\Xi$ process has been studied in several previous theoretical works employing various methodologies, including the effective Lagrangian approach~\cite{Sharov:2011xq, Shyam:2011ys, Jackson:2015dva}, the chiral unitary approach~\cite{Feijoo:2015yja}, a model-independent analysis~\cite{Jackson:2013bba}, coupled-channel models and partial-wave analyses~\cite{Kamano:2014zba, Matveev:2019igl}, the LASSO and BIC method~\cite{Landay:2018wgf}, and a hybrid Regge-plus-resonance approach~\cite{Kim:2023jij}. 
While these studies have achieved significant progress in describing the $K^-p \to K\Xi$ reactions, some observed structures, particularly the behavior near the threshold and the bump at $\sqrt{s}\approx2.3$ GeV in the $K^+\Xi^-$ channel, are not yet fully understood and call for further investigation. 
These features may indicate the possible presence of intermediate $S=-1$ hyperon resonances that are not yet accounted for in existing theoretical descriptions.

In this study, we employ an effective Lagrangian approach to estimate the $s$- and $u$-channel contributions to $K^-p\to K\Xi$ at tree level. 
Our model incorporates the $u$-channel $\Sigma$ and $\Lambda$ exchanges, $s$-channel $\Sigma$, $\Lambda$, $\Lambda(1800)$ and $\Sigma(2250)$ resonances, and we introduce a hypothetical $S = -1$ hyperon with $J^P = 3/2^-$ in the $s$-channel.
The latter two resonances significantly improve the description of the observed structures in the total cross sections, especially the enhancement near threshold and the bump near $\sqrt s \approx 2.3 $ GeV
observed exclusively in the $K^+\Xi^-$ channel.
The angular distribution strongly motivates the existence of an additional $J^P = 3/2^-$ state, requiring a dominant $D$-wave component. 
This is further supported by the rapid threshold rise of the total cross section, which also points to a sizeable $D$-wave contribution.
Furthermore, the 2024 Particle Data Group (PDG) compilation~\cite{ParticleDataGroup:2024cfk} lacks any $J^P=3/2^-$ $\Lambda^*$ resonance in the 1700--2000 MeV mass range, while the $\Sigma(1910)$ exhibits only a weak coupling to the $K\Xi$ channel.
This naturally leads to the question: where is the hyperon resonance that couples strongly to $s\bar{s}qqs$ configurations?
Confirming the presence of such a $J^P=3/2^-$ hyperon in existing experimental data would significantly inform future dedicated searches in this energy window.
Several predictions for the polarization of the final-state $\Xi$ and for the cross section of the $K^-n \to K^0\Xi^-$ reaction are also provided, which can be tested in future experiments at J-PARC~\cite{Aoki:2021cqa},  HIAF~\cite{Chen:2025ppt} and JLab~\cite{Dobbs:2022agy}.

The article is organized as follows.  Sec.~\ref{sec2} introduces our theoretical framework, Sec.~\ref{sec3} presents our results and discussion, and we conclude in Sec.~\ref{sec4} with a summary and some outlook.

\section{Formalism}\label{sec2}

\begin{figure}[tb]
	\centering
	\subfigure[]{
		\centering
		\includegraphics[width=4.5cm]{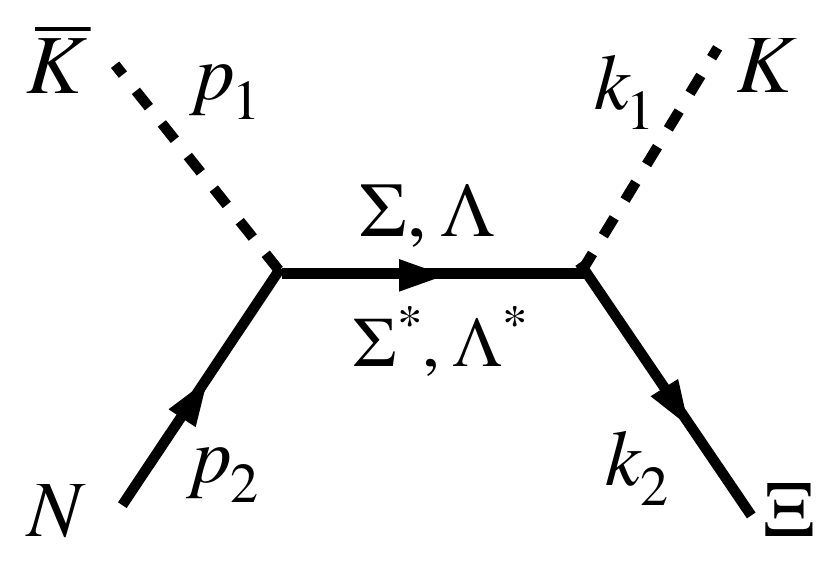}
	}
	\centering
	\subfigure[]{
		\centering
		\includegraphics[width=3.5cm]{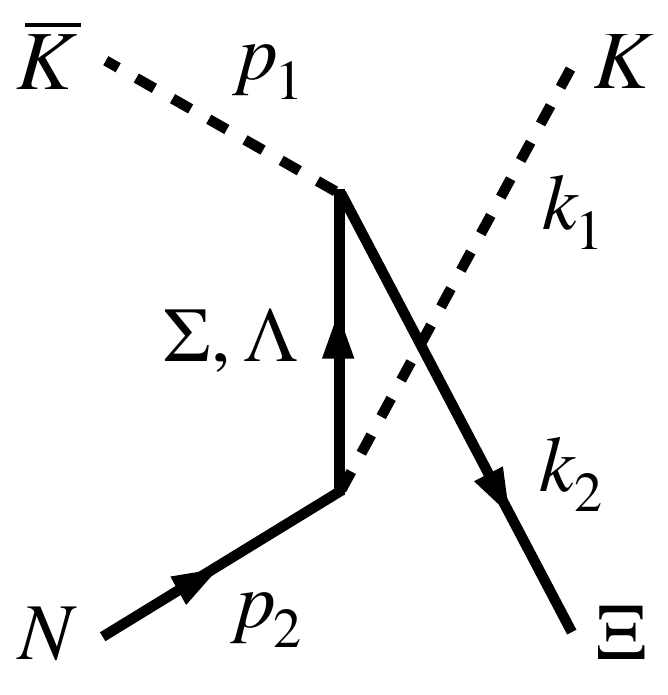}
	}
	\centering
	\caption{Tree-level Feynman diagrams for the reaction $\bar KN\to K\Xi$ with four-momenta labeled: (a) $s$-channel exchange of $\Sigma$, $\Lambda$, and their resonances; (b) $u$-channel exchange of $\Sigma$ and $\Lambda$. }
    \label{Feynman}
\end{figure}

In this section, we outline the effective Lagrangian approach employed to investigate the $K^-p\to K^+\Xi^-$ and $K^-p\to K^0\Xi^0$ reactions. The Feynman diagrams for the $s$- and $u$-channel processes are shown in Fig.~\ref{Feynman}, with the corresponding four-momenta labeled. The $u$-channel contributions from $\Sigma^*$ and $\Lambda^*$ resonances are negligible compared to those from the ground states, since their larger masses strongly suppress the propagators. The $t$-channel contributions are absent due to the lack of $S = 2$ mesons.
\subsection{Effective Lagrangians and form factors}
The effective Lagrangians for ground-state $\Sigma$ and $\Lambda$ exchanges are given as follows: 
\begin{equation}
    \mathcal L_{\Sigma N\bar K}=g_{\Sigma N\bar K}\,\bar{N}\bm{\tau} (D_{N\Sigma }K)\cdot \bm{\Sigma}+\text{h.c.},
\end{equation}
\begin{equation}
    \mathcal L_{\Sigma \Xi K}=g_{\Sigma \Xi K}\,\bar{\Xi}\bm{\tau} (D_{\Xi\Sigma }K_c)\cdot \bm{\Sigma}+\text{h.c.},
\end{equation}
\begin{equation}
    \mathcal L_{\Lambda N\bar K}=g_{\Lambda N\bar K}\,\bar{N} (D_{N\Lambda }K)\Lambda+\text{h.c.},
\end{equation}
\begin{equation}
    \mathcal L_{\Lambda \Xi K}=g_{\Lambda \Xi K}\,\bar{\Xi}(D_{\Xi\Lambda }K_c)\Lambda+\text{h.c.},
\end{equation}
where the operator $D_{B'B}$ is defined as
\begin{equation}
    D_{B'B}=-\gamma^5\left(i\lambda+\frac{1-\lambda}{m_{B}+m_{B'}}\slashed\partial\right),
\end{equation}
with $\lambda = 0$ and $\lambda = 1$ corresponding to the pseudovector and pseudoscalar coupling schemes, respectively. The notations of isospin multiplets are defined in Appendix~\ref{app:1}, and the values of coupling constants and the parameter $\lambda$ are listed in Table~\ref{tab:1}.

For the $S = -1$ resonances considered in this work, the effective Lagrangians are constructed as follows:
\begin{equation}
    \mathcal L_{\Lambda^*(1/2^-) N\bar K}=-ig_{\Lambda^*(1/2^-) N\bar K}\,\bar{N} K{\Lambda}^*+\text{h.c.},
\end{equation}
\begin{equation}
    \mathcal L_{\Lambda^*(1/2^-) \Xi K}=-ig_{\Lambda^*(1/2^-) \Xi K}\,\bar{\Xi} K_c{\Lambda}^*+\text{h.c.},
\end{equation}
\begin{equation}
    \mathcal L_{\Lambda^*(3/2^-) N\bar K}=\frac{g_{\Lambda^*(3/2^-)N\bar K}}{m_K} \,\bar{N}\gamma^5 (\partial_{\mu}K)\Lambda^{*\mu}+\text{h.c.},
\end{equation}
\begin{equation}
    \mathcal L_{\Lambda^*(3/2^-) \Xi K}=\frac{g_{\Lambda^*(3/2^-)\Xi K}}{m_K} \,\bar{\Xi} \gamma^5(\partial_{\mu}K_c)\Lambda^{*\mu}+\text{h.c.},
\end{equation}
\begin{equation}
    \mathcal L_{\Sigma^*(3/2^-) N\bar K}=\frac{g_{\Sigma^*(3/2^-)N\bar K}}{m_K} \,\bar{N}\gamma^5\bm{\tau} (\partial_{\mu}K)\cdot \bm{\Sigma}^{*\mu}+\text{h.c.},
\end{equation}
\begin{equation}
    \mathcal L_{\Sigma^*(3/2^-) \Xi K}=\frac{g_{\Sigma^*(3/2^-)\Xi K}}{m_K} \,\bar{\Xi}\gamma^5\bm{\tau} (\partial_{\mu}K_c)\cdot \bm{\Sigma}^{*\mu}+\text{h.c.},
\end{equation}
\begin{equation}
    \mathcal L_{\Sigma^*(5/2^-) N\bar K}=\frac{ig_{\Sigma^*(5/2^-)N\bar K}}{m_K^2} \,\bar{N}\bm{\tau} (\partial_{\mu}\partial_{\nu}K)\cdot \bm{\Sigma}^{*\mu\nu}+\text{h.c.},
\end{equation}
\begin{equation}
    \mathcal L_{\Sigma^*(5/2^-) \Xi K}=\frac{ig_{\Sigma^*(5/2^-)\Xi K}}{m_K^2} \,\bar{\Xi}\bm{\tau} (\partial_{\mu}\partial_{\nu}K_c)\cdot \bm{\Sigma}^{*\mu\nu}+\text{h.c.}.
\end{equation}
The $J^P$ labels in parentheses denote the quantum numbers of the corresponding resonances. The values of the related coupling constants are listed in Table~\ref{tab:1}.

To account for the finite size of hadrons, we introduce a hadronic form factor at each vertex:
\begin{equation}
    F_B(q^2)=\frac{\Lambda_B^4}{(q^2-m_B^2)^2+\Lambda_B^4},
\end{equation}
where $q$ and $m_B$ are the four-momentum and mass of the exchanged baryon $B$, respectively. The cutoff parameter $\Lambda_B$ characterizes the typical hadronic scale, and its specific value for each exchanged baryon is listed in Table~\ref{tab:1}.
\subsection{Propagators}
Higher-spin fields suffer from Johnson--Sudarshan~\cite{Johnson:1960vt} and Velo--Zwanziger~\cite{Velo:1969bt,Velo:1969txo} problems, which originate from the violation of constraints in the presence of interactions~\cite{Hasumi:1979db,Cox:1989hp,Pascalutsa:1998pw}. These issues cause unphysical lower-spin components to mix into the propagation of higher-spin particles off shell, thereby violating unitarity. For instance, in the $s$-channel process $\bar KN\to K\Xi$ with a $\Sigma^*(3/2^-)$ exchanged, both $\Sigma^*(3/2^-)N\bar K$ and $\Sigma^*(3/2^-)\Xi K$ vertices are in pure $D$-wave, for which the differential cross section is expected to exhibit forward-backward symmetry. If the standard Rarita-Schwinger propagator~\cite{Rarita:1941mf}:
\begin{equation}
    S^{\mathrm{std}}_{\mu\nu}(q)=\frac{\slashed q+m}{q^2-m^2+im\Gamma}\left(-g_{\mu\nu}+\frac{1}{3}\gamma_\mu\gamma_\nu+\frac{\gamma_\mu q_\nu-\gamma_\nu q_\mu}{3m}+\frac{2q_\mu q_\nu}{3m^2}\right)
\end{equation}
is used to describe the propagation of the $\Sigma^*(3/2^-)$, the unphysical spin-$\frac{1}{2}$ component will be activated when this intermediate particle is off shell, since $\gamma^\mu S^{\mathrm{std}}_{\mu\nu}(q)$ and $q^\mu S^{\mathrm{std}}_{\mu\nu}(q)$ do not vanish. This leads to a mixing of $P$- and $D$-waves at each vertex, breaking the forward-backward symmetry of the differential cross section. Consequently, the standard Rarita–Schwinger propagator violates both the unitarity and the purity of the partial-wave decomposition. 

One way to resolve this issue is to employ gauge-invariant Lagrangians~\cite{Pascalutsa:1998pw} to protect unitarity. Alternatively, for phenomenological calculations with the conventional couplings adopted here, a consistent description can be achieved by replacing the higher-spin propagators with off-shell projection operators~\cite{Adelseck:1985scp,Benmerrouche:1989uc,Pascalutsa:1998pw}. 

The explicit forms of the propagators employed in this work are listed below. Propagators for spin-$\frac{1}{2}$ baryons have the standard form:
\begin{equation}
    S(q)=\frac{\slashed q+m}{q^2-m^2+im\Gamma},
\end{equation}
$q$, $m$ and $\Gamma$ are the four-momentum, mass and width of the exchanged particle, respectively. 
Propagators for spin-$\frac{3}{2}$ baryons take the form of the off-shell projection operators:
\begin{equation}
    S_{\mu\nu}(q)=\frac{\slashed q+\sqrt{q^2}}{q^2-m^2+im\Gamma}\Delta_{\mu\nu}(q),
\end{equation}
with
\begin{equation}
    \Delta_{\mu\nu}=-\tilde g_{\mu\nu}+\frac{1}{3}\tilde\gamma_\mu\tilde\gamma_\nu,
\end{equation}
where $\tilde g_{\mu\nu}=g_{\mu\nu}-q^\mu q^\nu/q^2$ and $\tilde\gamma_\mu=\tilde g_{\mu\nu}\gamma^\nu$. One can verify that $q^\mu S_{\mu\nu}(q)$ and $\gamma^\mu S_{\mu\nu}(q)$ vanish, indicating that $S_{\mu\nu}(q)$ propagates purely spin-$\frac{3}{2}$ components. Analogously, propagators for spin-$\frac{5}{2}$ baryons take the form:
\begin{equation}
    S_{\mu\nu,\alpha\beta}(q)=\frac{\slashed q+\sqrt{q^2}}{q^2-m^2+im\Gamma}\Delta_{\mu\nu,\alpha\beta}(q),
\end{equation}
with
\begin{equation}
    \Delta_{\mu\nu,\alpha\beta}=\frac{1}{2}(\tilde g_{\mu\alpha}\tilde g_{\nu\beta}+\tilde g_{\mu\beta}\tilde g_{\nu\alpha})-\frac{1}{5}\tilde g_{\mu\nu}\tilde g_{\alpha\beta}-\frac{1}{10}(\tilde g_{\mu\alpha}\tilde \gamma_\nu\tilde\gamma_\beta+\tilde g_{\mu\beta}\tilde \gamma_\nu\tilde\gamma_\alpha+\tilde g_{\nu\alpha}\tilde \gamma_\mu\tilde\gamma_\beta+\tilde g_{\nu\beta}\tilde \gamma_\mu\tilde\gamma_\alpha),
\end{equation}
in which $\mu\nu$ and $\alpha\beta$ correspond to the final and initial indices.
\subsection{Amplitudes and cross section}
The total amplitudes for the $\bar KN\to K\Xi$ reactions are constructed from the Feynman diagrams shown in Fig.~\ref{Feynman}, 
\begin{equation}\label{eq:21}
    \mathcal M_{K^-p\to K^+\Xi^-}=-\mathcal M^s_\Sigma-\mathcal M^u_\Sigma+\mathcal M^s_\Lambda+\mathcal M^u_\Lambda-\sum_{\Sigma^*}e^{i\phi_{\Sigma^*}}\mathcal M^s_{\Sigma^*}+\sum_{\Lambda^*}e^{i\phi_{\Lambda^*}}\mathcal M^s_{\Lambda^*},
\end{equation}
\begin{equation}\label{eq:22}
    \mathcal M_{K^-p\to K^0\Xi^0}=\mathcal M^s_\Sigma-2\mathcal M^u_\Sigma+\mathcal M^s_\Lambda+\sum_{\Sigma^*}e^{i\phi_{\Sigma^*}}\mathcal M^s_{\Sigma^*}+\sum_{\Lambda^*}e^{i\phi_{\Lambda^*}}\mathcal M^s_{\Lambda^*},
\end{equation}
\begin{equation}
    \mathcal M_{K^-n\to K^0\Xi^-}=\mathcal M_{K^-p\to K^+\Xi^-}-\mathcal M_{K^-p\to K^0\Xi^0}.
\end{equation}
Here, $\mathcal M_X^{s(u)}$ denotes the amplitude of the $s$-channel ($u$-channel) exchange of particle $X$, and the numerical coefficients in front of each term are the corresponding isospin factors. The explicit expressions of all amplitudes are provided in Appendix~\ref{app:2}. The phase factors $e^{i\phi_{\Sigma^*}}$ and $e^{i\phi_{\Lambda^*}}$ account for possible loop effects at the vertices involving $\Sigma^*$ and $\Lambda^*$ exchange\footnote{Under SU(3) flavor symmetry, the phase factors associated with the ground-states $\Sigma$ and $\Lambda$ exchanges are identical, thus can be absorbed into an overall phase of the total amplitude and therefore do not affect the physical observables. For this reason, they are omitted in Eqs.~\eqref{eq:21} and~\eqref{eq:22}.}, respectively, and are treated as free parameters in our fitting procedure. The summations in Eqs.~\eqref{eq:21} and~\eqref{eq:22} run over all resonances included in our model. The resonance parameters, along with the fitted values of their associated phases, are listed in Tables~\ref{tab:1} and \ref{tab:2}. 

The differential cross section for the $K^-p\to K^+\Xi^-$ reaction in the center-of-mass (c.m.) frame is given by
\begin{equation}
    \frac{\text{d}\sigma_{K^-p\to K^+\Xi^-}}{\text{d}\Omega}=\frac{1}{64\pi^2 s}\frac{|\bm{k}_1|}{|\bm{p}_1|}\cdot\frac{1}{2}\sum_{r_2,s_2}\left|\mathcal M_{K^-p\to K^+\Xi^-}(r_2,s_2)\right|^2,
\end{equation}
where $\bm{p}_1$ and $\bm{k}_1$ denote the 3-momenta of the initial $K^-$ and the final $K^+$ in the c.m. frame, respectively, and $s=(p_1+p_2)^2$. The $s_2$ and $r_2$ denote the spin indices of the initial proton and the final $\Xi^-$, respectively, and the factor $1/2$ accounts for the spin average over the initial proton. The differential cross sections for $K^-p\to K^0\Xi^0$ and $K^-n\to K^0\Xi^-$ can be written similarly.

We also calculate the polarization of the final-state $\Xi$ in the reaction $K^-p\to K\Xi$ for an unpolarized initial state. Taking the reaction plane as the $xz$-plane, the polarization of the $\Xi^-$ along the $y$-axis is given by
\begin{equation}
    P_{\Xi^-}=\frac{-2\sum_{s_2}{\rm Im}\mathcal M_{K^-p\to K^+\Xi^-}(\uparrow,s_2)\mathcal M^*_{K^-p\to K^+\Xi^-}(\downarrow,s_2)}{\sum_{r_2,s_2}\left|\mathcal M_{K^-p\to K^+\Xi^-}(r_2,s_2)\right|^2}.
\end{equation}
Here, $\uparrow$ and $\downarrow$ denote the spin of the $\Xi^-$ along the $z$-axis. The polarization of the $\Xi^0$ in the reaction $K^-p\to K^0\Xi^0$ can be written analogously.

\section{Results and discussion}\label{sec3}

We first discuss the physical motivation for the set of particles included in our model. 
The experimental differential cross sections exhibit a pronounced backward enhancement, indicating that $u$-channel exchanges of the ground states $\Sigma$ and $\Lambda$ play an important role. 
The total cross sections for both $K^+\Xi^-$ and $K^0\Xi^0$ channels rise rapidly above threshold and form a prominent structure around 2.0~GeV, and such behavior may indicate the presence of an intermediate resonance that couples to the initial and final states in higher partial waves. 
In particular, this feature is consistent with the contributions from a $J^P = 3/2^-$ hyperon, which could correspond to either a $\Lambda^*(3/2^-)$ or a $\Sigma^*(3/2^-)$ state. 
The bump around 2.3~GeV in the total cross section can be associated with the $\Sigma(2250)$ resonance. 
According to the partial-wave analysis in Ref.~\cite{deBellefon:1976qr}, the $\Sigma(2250)$ has been assigned $J^P = 5/2^-$ or $9/2^-$. 
To avoid introducing an excessively high partial wave which would tend to shift the bump to higher energies, we adopt the $J^P = 5/2^-$ assignment. 
Finally, the $\Lambda(1800)$ resonance is included to provide additional interference near threshold, 
which improves the description of the total cross sections in this region. 

We perform a fit of our model to the available total and differential cross section data for both $K^+\Xi^-$ and $K^0\Xi^0$ channels, including 386 data points and 9 parameters. 
Polarization data of $\Xi$ and the cross section data of $K^-n\to K^0\Xi^-$ are not included in the fitting procedure.
Given the large uncertainties in the current data, as well as total and differential cross sections, the polarization and $K^-n\to K^0\Xi^-$ data are served as independent validation to distinguish our two models. 

The $\chi^2$ function is defined as
\begin{equation}
    \chi^2 = \sum_{i} \frac{\left( \mathcal{O}_i^{\text{exp}} - \mathcal{O}_i^{\text{th}} \right)^2}{\sigma_i^2},
\end{equation}
where $\mathcal{O}_i^{\text{exp}}$ and $\mathcal{O}_i^{\text{th}}$ denote the experimental and theoretical values of the $i$-th data point, respectively, and $\sigma_i$ is the corresponding uncertainty\footnote{For experimental measurements without reported errors, the statistical uncertainty is estimated as $\sqrt{N}$, where $N$ is the number of events.}. 
In practice, some data points are reported with very small statistical uncertainties, which can overly constrain the fit and lead to distortions in the overall behavior of the theoretical curves. 
To mitigate this issue, a moderate weighting scheme is adopted to balance the contributions from different data points. 
We have verified that the main features of the fit remain stable under reasonable variations of the weighting procedure.

The available data can be described by two distinct fitting solutions. 
In Solution-I, the observed structures around 2~GeV are attributed to a $\Sigma^*(3/2^-)$ resonance, while in Solution-II, they are instead described by a $\Lambda^*(3/2^-)$ resonance. 
Both scenarios provide comparable descriptions of the existing data, and cannot be distinguished at the present level of experimental precision. 
In the following, we present and compare the results of the two solutions for each observable, highlighting their similarities and differences.

\subsection{Fitted parameters and overall fit quality}

\begin{table}[tb]
    \centering
    \caption{Particles and parameters involved in Solution-I. The $\Sigma^*$ with $J^P=\frac{3}{2}^-$ denotes the hypothetical hyperon introduced in this fitting. The masses and widths of established states are taken from the Particle Data Group~\cite{ParticleDataGroup:2024cfk}, while the coupling constants $g_{\Sigma N\bar K}$, $g_{\Sigma \Xi K}$, $g_{\Lambda N\bar K}$ and $g_{\Lambda \Xi K}$ are determined using SU(3) flavor symmetry relations~\cite{Ronchen:2012eg}. The cutoff parameters $\Lambda_B$ are fixed in the fitting procedure. All remaining parameters, indicated by a dagger, are treated as free parameters and determined by fitting to the data. }
    \begin{tabular}{lcccccccc}
    \hline
    Baryons & {\quad$J^P$\quad} & {\quad$m_B$[MeV]\quad} & {\quad$\Gamma_B$[MeV]\quad}  & {\quad$g_{BN\bar K}g_{B\Xi K}$\quad} & {\quad$\lambda$\quad} & {\quad$\phi_B$\quad} & {\quad$\Lambda_B$[GeV]\quad} & Rating\\
    \hline
    $\Lambda$ & $\frac{1}{2}^+$ & $1116$ & $0$ & $-13.98\times4.66 $ & \multirow{2}{*}{$0.38\,^\dagger$} & - &  1.23 & **** \\
    $\Sigma$ & $\frac{1}{2}^+$ & $1193$ & $0$ & $2.69\times-13.45 $&  & - &  0.80 & ****\\
    $\Lambda(1800)$ & $\frac{1}{2}^-$ & $1800$ & $200$ & $-0.2\,^\dagger$ & - & $-80^\circ\,^\dagger$ &  1.30 & ***\\
    $\Sigma^*$ & $\frac{3}{2}^-$ & $1890\,^\dagger$ & $240\,^\dagger$ & $-30\,^\dagger$& - & $-83^\circ\,^\dagger$ &  1.30 & -\\
    $\Sigma(2250)$ & $\frac{5}{2}^-$ & $2270$ & $120$ & $0.06\,^\dagger$& - & $57^\circ\,^\dagger$ &  1.30 & **\\
    \hline
    \end{tabular}
    \label{tab:1}
\end{table}

\begin{table}[tb]
    \centering
    \caption{Particles and parameters involved in Solution-II. The $\Lambda^*$ with $J^P=\frac{3}{2}^-$ denotes the hypothetical hyperon introduced in this fitting. The fixed parameters are the same as Table~\ref{tab:1}, and the free parameters which are indicated by a dagger, are determined by fitting to the data. }
    \begin{tabular}{lcccccccc}
    \hline
    Baryons & {\quad$J^P$\quad} & {\quad$m_B$[MeV]\quad} & {\quad$\Gamma_B$[MeV]\quad}  & {\quad$g_{BN\bar K}g_{B\Xi K}$\quad} & {\quad$\lambda$\quad} & {\quad$\phi_B$\quad} & {\quad$\Lambda_B$[GeV]\quad} & Rating\\
    \hline
    $\Lambda$ & $\frac{1}{2}^+$ & $1116$ & $0$ & $-13.98\times4.66 $ & \multirow{2}{*}{$0.38\,^\dagger$} & - &  1.23 & **** \\
    $\Sigma$ & $\frac{1}{2}^+$ & $1193$ & $0$ & $2.69\times-13.45 $&  & - &  0.80 & ****\\
    $\Lambda(1800)$ & $\frac{1}{2}^-$ & $1800$ & $200$ & $-0.2\,^\dagger$ & - & $-77^\circ\,^\dagger$ &  1.30 & ***\\
    $\Lambda^*$ & $\frac{3}{2}^-$ & $1890\,^\dagger$ & $220\,^\dagger$ & $30\,^\dagger$& - & $-77^\circ\,^\dagger$ &  1.30 & -\\
    $\Sigma(2250)$ & $\frac{5}{2}^-$ & $2270$ & $120$ & $0.06\,^\dagger$& - & $40^\circ\,^\dagger$ &  1.30 & **\\
    \hline
    \end{tabular}
    \label{tab:2}
\end{table}

The fitted parameters for both solutions are summarized in Tables~\ref{tab:1} and~\ref{tab:2}. 
Solution-I yields $\chi^2/\mathrm{d.o.f.} = 3.05$, while Solution-II gives $\chi^2/\mathrm{d.o.f.}=3.20$, indicating that both scenarios provide comparable and satisfactory descriptions of the available data.

In both solutions, the mass of the introduced $J^P=3/2^-$ state is fitted to be around $1890$~MeV, suggesting that the structure around 2.0~GeV in the total cross sections is robustly associated with a $J^P=3/2^-$ hyperon, regardless of its isospin assignment. 
The slightly different widths ($240$~MeV for $\Sigma^*(3/2^-)$ vs.\ $220$~MeV for $\Lambda^*(3/2^-)$) reflect the distinct interference patterns arising from the isospin structures of the two scenarios.

The most notable difference between the two solutions lies in the distinct relative phases between $\Sigma^*(3/2^-)$ and $\Lambda^*(3/2^-)$ in  Eqs.~\eqref{eq:21} and~\eqref{eq:22}. 
While the total cross sections and $K^+\Xi^-$ channel differential cross section remain similar in the two solutions, the differing relative phases lead to visible different interferences in $K^0\Xi^0$ channel differential cross section, especially at $\sqrt{s}= 2.28$~GeV due to different phases of $\Sigma(2250)$ in the two solutions. 
Let us discuss the total and differential cross section in detail in the next two subsections.

At last, it should be noted that the $\Sigma^*(3/2^-)$ resonance introduced in Solution-I is not to be identified with the $\Sigma(1910)3/2^-$ listed in the PDG. 
We have explicitly checked that the $\Sigma(1910)$ has very small couplings to the $K^-p$ and $K\Xi$ channels and therefore cannot provide a significant contribution to the structure around 2.0~GeV in the present reaction.

\subsection{Total cross sections}

\begin{figure}[tb]
    \centering
    \begin{subfigure}
        \centering
        \includegraphics[width=0.48\linewidth]{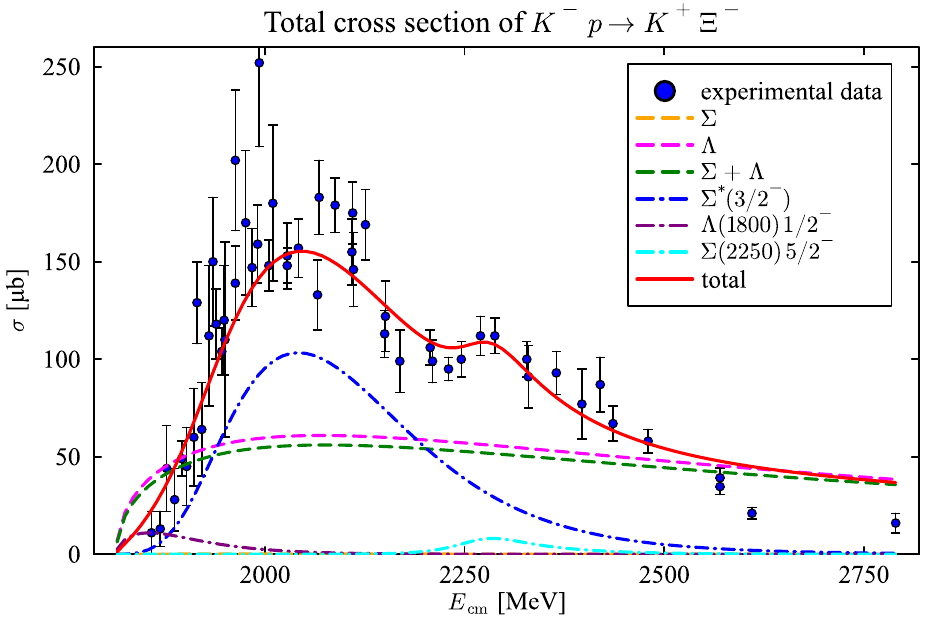}
    \end{subfigure}
    \hfill
    \begin{subfigure}
        \centering
        \includegraphics[width=0.48\linewidth]{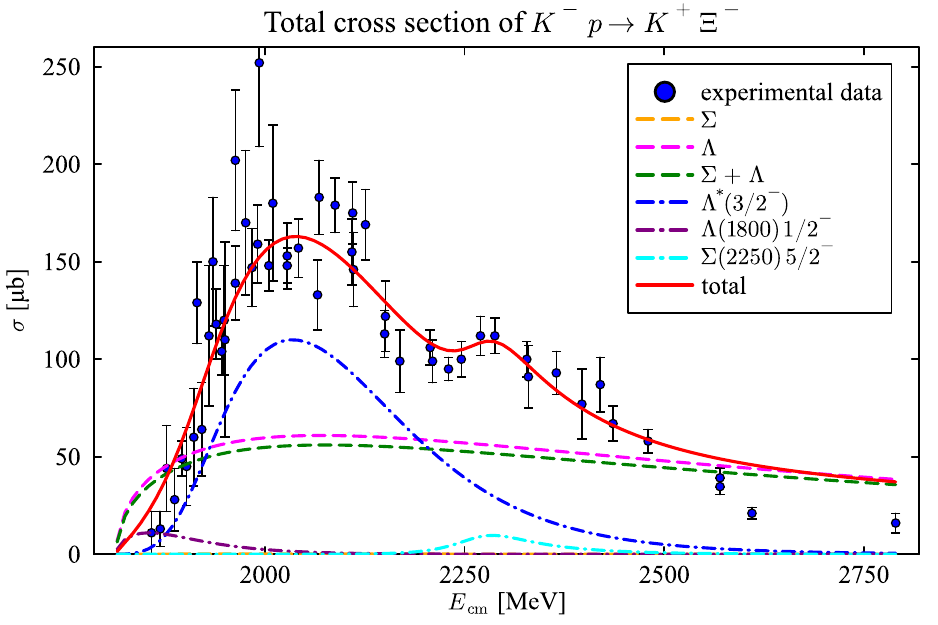}
    \end{subfigure}
    \caption{Total cross sections of $K^-p\to K^+\Xi^-$ in Solution-I (left) and Solution-II (right). The experimental data are taken from Refs.~\cite{Cooper:1962rxa,Carmony:1964zza,Badier:1965zzc,Berge:1966zz,Birmingham-Glasgow-LondonIC-Oxford-Rutherford:1966onr,London:1966zz,Trippe:1967wat,Burgun:1968ice,Trower:1968zz,Dauber:1969hg,Griselin:1975pa,Briefel:1977bp,deBellefon:1977sw}.
    The curve of $\Sigma$ is invisible because of its small contribution.}
    \label{fig:TC_charged}
\end{figure}

\begin{figure}[tb]
    \centering
    \begin{subfigure}
        \centering
        \includegraphics[width=0.48\linewidth]{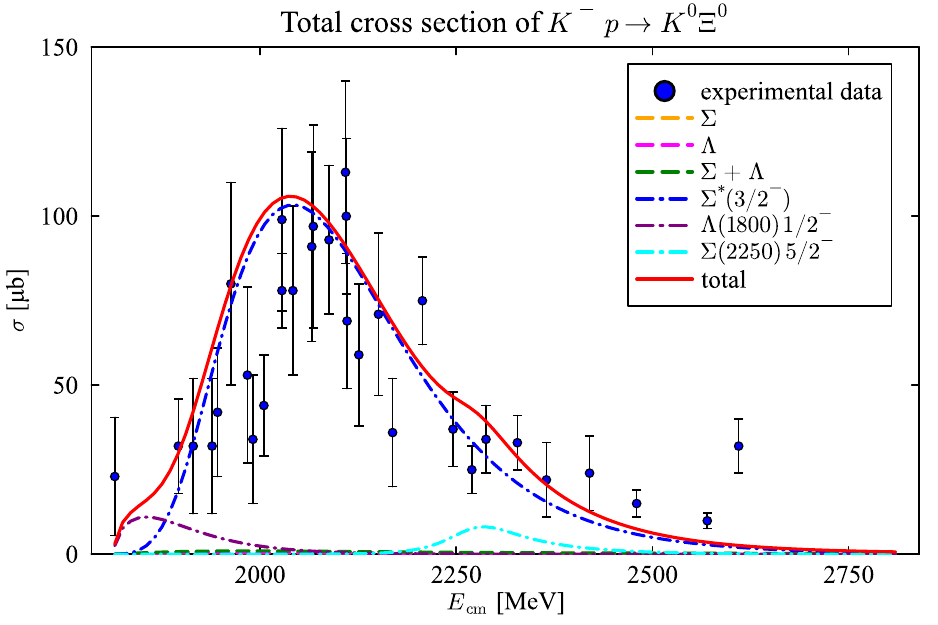}
    \end{subfigure}
    \hfill
    \begin{subfigure}
        \centering
        \includegraphics[width=0.48\linewidth]{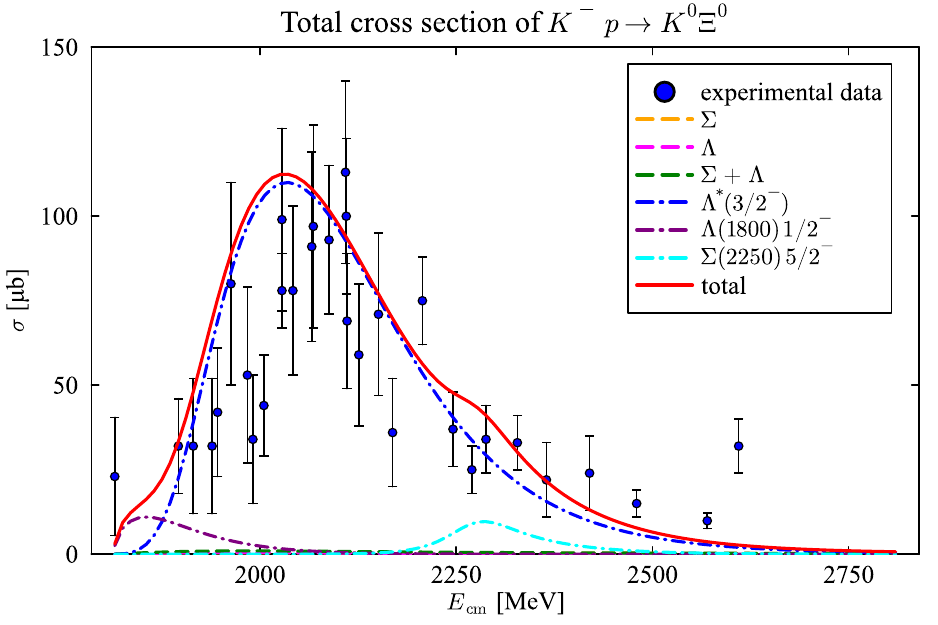}
    \end{subfigure}
    \caption{Total cross sections of $K^-p\to K^0\Xi^0$ in Solution-I (left) and Solution-II (right). The experimental data are taken from Refs.~\cite{Badier:1965zzc,Berge:1966zz,Burgun:1968ice,Dauber:1969hg,Carlson:1973td,Briefel:1977bp,deBellefon:1977sw}. The curves of $\Sigma$ and $\Lambda$ are invisible because of their small contribution.}
    \label{fig:TC_neutral}
\end{figure}

The total cross sections for the $K^+\Xi^-$ and $K^0\Xi^0$ channels are shown in Figs.~\ref{fig:TC_charged} and~\ref{fig:TC_neutral}, respectively.
Both solutions provide a satisfactory overall description of the experimental data across the entire energy range $1.8 < \sqrt{s} < 2.8$~GeV.

The most prominent feature in both channels is the rapid rise of the cross section near threshold, followed by a broad structure peaking around $\sqrt{s} \approx 2.0$~GeV. 
In both solutions, this behavior is dominantly driven by the $s$-channel exchange of the $J^P=3/2^-$ state. 
The $D$-wave nature of the couplings-specifically, the derivative vertices in the effective Lagrangians for the spin-$3/2$ resonances exchange (see Sec.~\ref{sec2}) makes the amplitude proportional to the initial and final kaon momenta near threshold,
naturally producing the observed rapid rise.
This kinematic behavior is thus well described by the $D$-wave $3/2^-$ resonance and provides strong motivation for its introduction.

In the $K^+\Xi^-$ channel, both solutions reproduce the bump around $\sqrt{s} \approx 2.3$~GeV, which is generated by the constructive interference between the $\Sigma(2250)$ resonance and other mechanisms. 
This bump is not obvious in the data of $K^0\Xi^0$ channel, while theoretical side, we still can find a slight bump.  
This phenomenon originates in that the $u$-channel exchange of $\Lambda$ is absent in the neutral channel final states, and the interference between $\Sigma(2250)$ and other mechanisms such as the $3/2^-$ state is not strong.

At last, for the higher energy region ($E_{\text{cm}}>2.5$ GeV), the $u$-channel exchange dominates the $K^+\Xi^-$ channel, while it almost drops to zero in the $K^0\Xi^0$ channel due to the absence of $u$-channel $\Lambda$ exchange.

\subsection{Differential cross sections}

\begin{figure}[tb]
    \centering
    \begin{subfigure}
        \centering
        \includegraphics[width=0.48\linewidth]{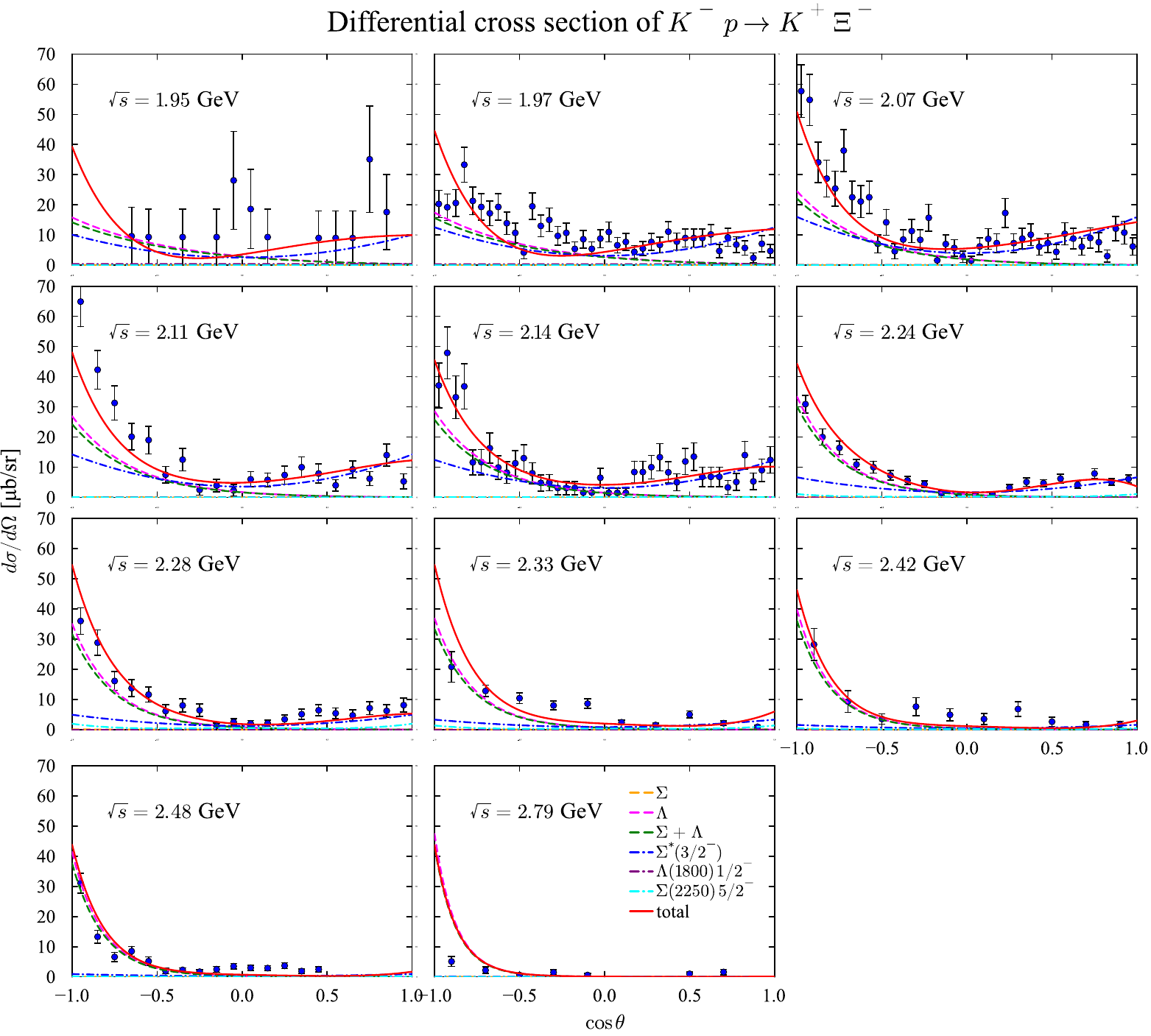}
    \end{subfigure}
    \hfill
    \begin{subfigure}
        \centering
        \includegraphics[width=0.48\linewidth]{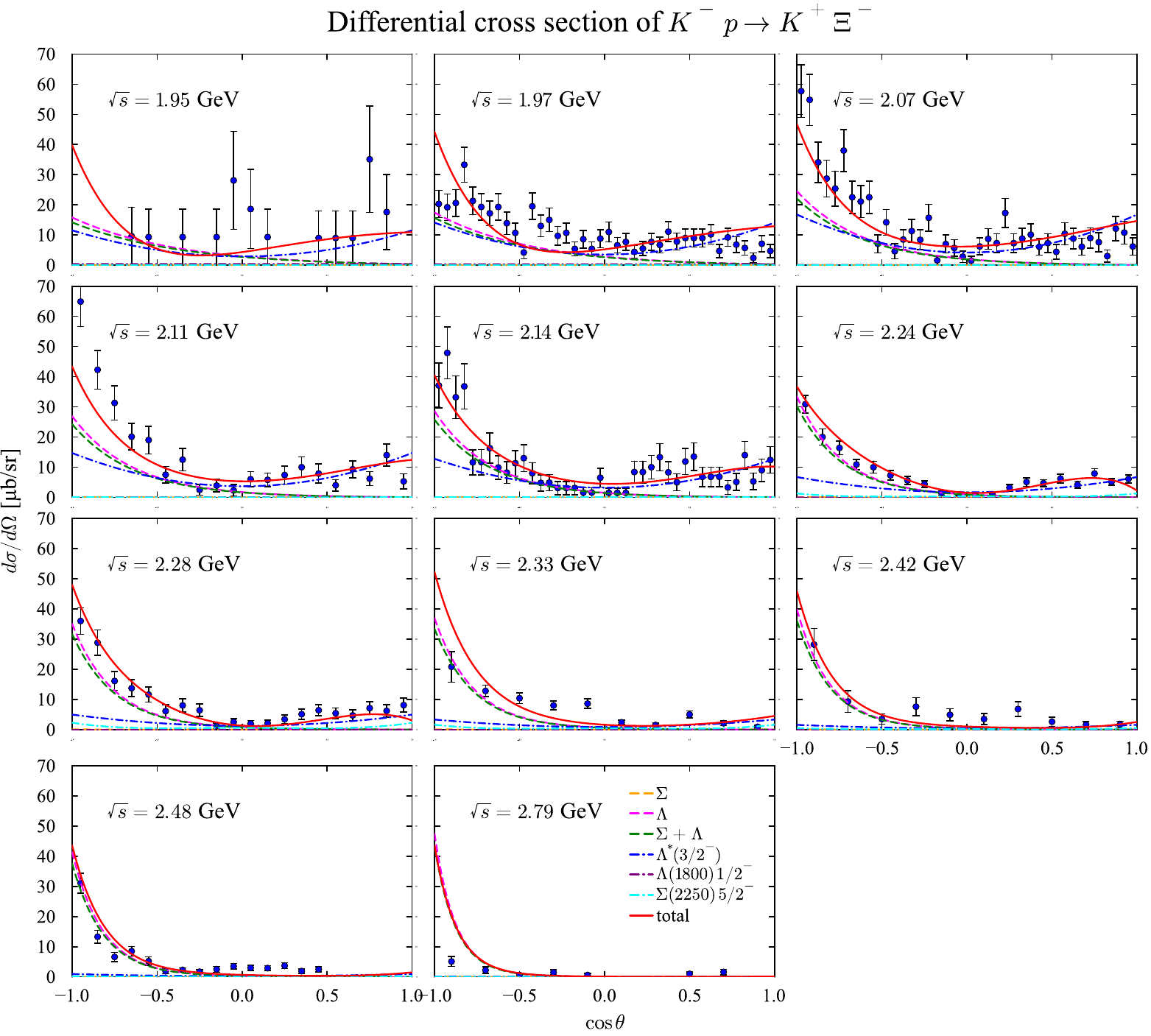}
    \end{subfigure}
    \caption{Differential cross sections of $K^-p\to K^+\Xi^-$ in Solution-I (left) and Solution-II (right) at various c.m.\ energies. The scattering angle $\theta$ is defined as the angle between the direction of the incoming $K^-$ and the outgoing $K^+$ in the c.m.\ frame. The experimental data are taken from Refs.~\cite{Birmingham-Glasgow-LondonIC-Oxford-Rutherford:1966onr,London:1966zz,Trippe:1967wat,Burgun:1968ice,Trower:1968zz,Dauber:1969hg}.}
    \label{fig:DSC_charged}
\end{figure}

\begin{figure}[tb]
    \centering
    \begin{subfigure}
        \centering
        \includegraphics[width=0.48\linewidth]{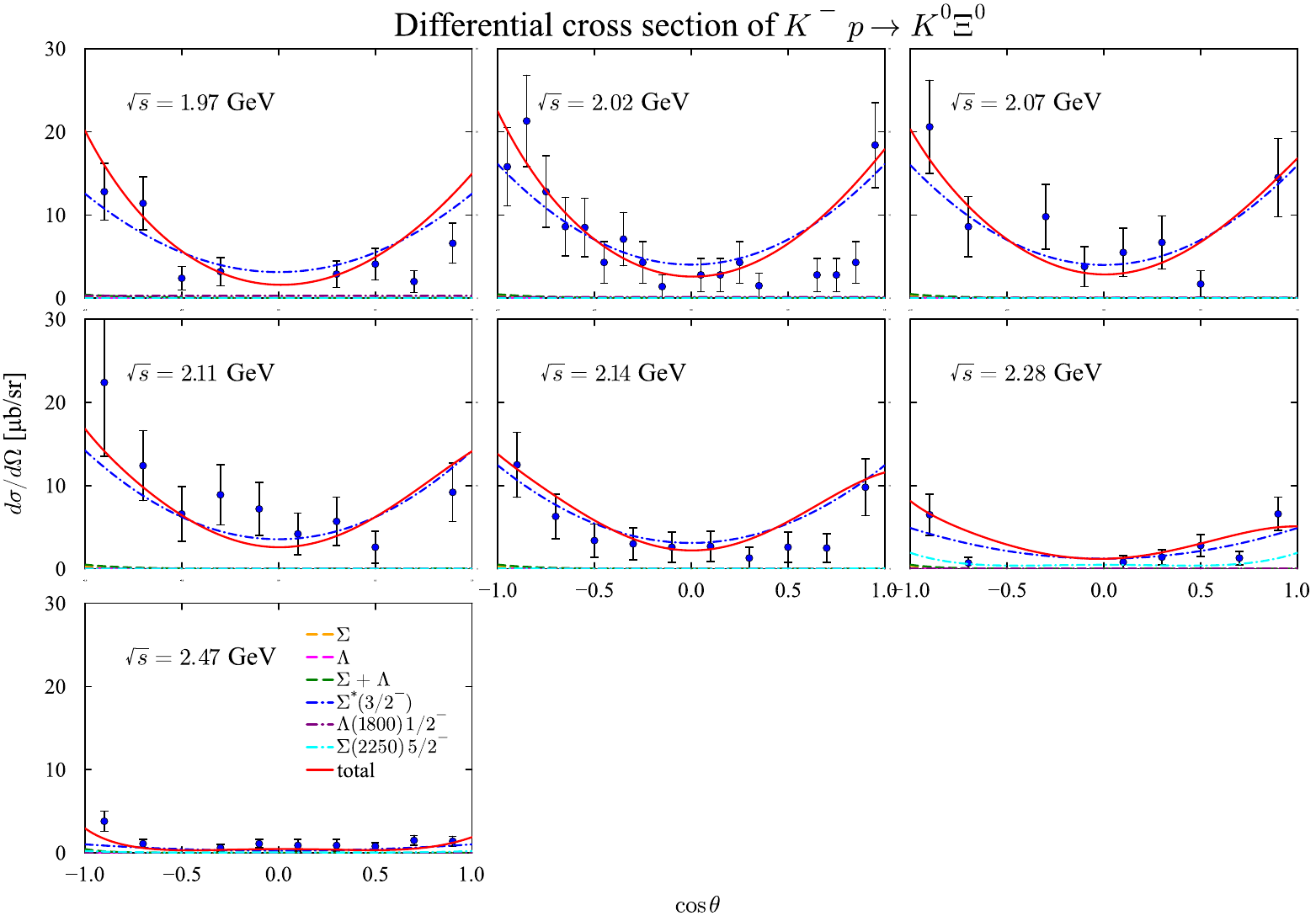}
    \end{subfigure}
    \hfill
    \begin{subfigure}
        \centering
        \includegraphics[width=0.48\linewidth]{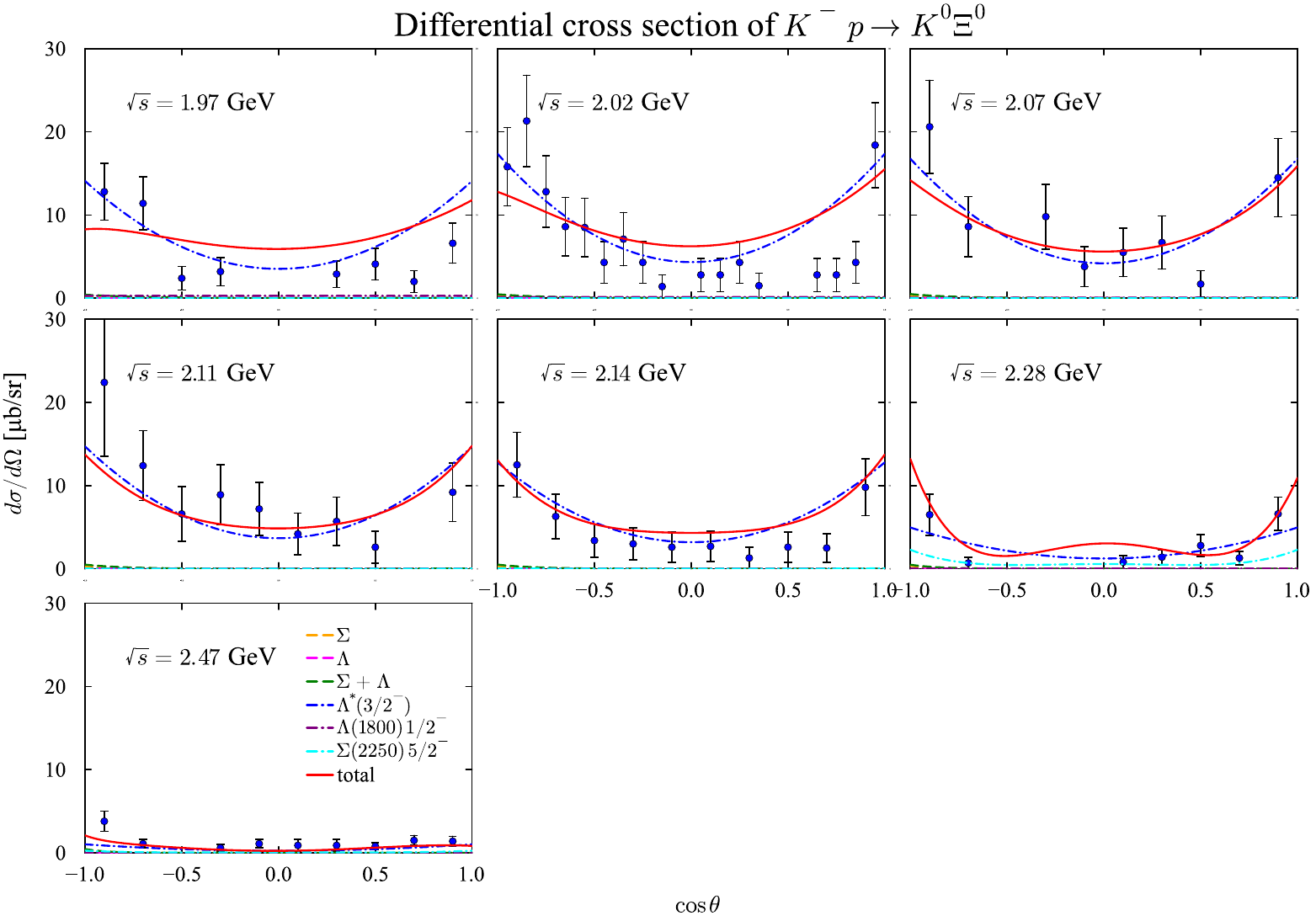}
    \end{subfigure}
    \caption{Differential cross sections of $K^-p\to K^0\Xi^0$ in Solution-I (left) and Solution-II (right) at various c.m.\ energies. The scattering angle $\theta$ is defined as the angle between the direction of the incoming $K^-$ and the outgoing $K^0$ in the c.m.\ frame. The experimental data are taken from Refs.~\cite{Berge:1966zz,Burgun:1968ice,Dauber:1969hg}.}
    \label{fig:DSC_neutral}
\end{figure}

The differential cross sections for both channels are presented in Fig.~\ref{fig:DSC_charged} and Fig.~\ref{fig:DSC_neutral}. 
The $K^+\Xi^-$ channel exhibits a pronounced backward enhancement ($\cos\theta \to -1$), and in our model, this backward peak is produced by the $u$-channel exchanges of the ground-state $\Sigma$ and $\Lambda$ hyperons. 
Both solutions describe this feature equally well, as the contributions of ground-state $\Sigma$ and $\Lambda$ are identical in the two scenarios, the only differences appear in the $s$-channel resonance sector.

In contrast, the $K^0\Xi^0$ differential cross sections are considerably flatter and nearly forward-backward symmetric.
This difference arises from that the $u$-channel $\Lambda$ exchange is absent in the $K^0\Xi^0$ channel (see Eq.~\eqref{eq:22}), which reduces the backward enhancement. 
The remaining $s$-channel contributions, dominated by the $D$-wave $3/2^-$ resonance, produce a more symmetric angular distribution.

The differential cross sections at low energies ($\sqrt{s} \lesssim 2.1$~GeV) are particularly sensitive to the $3/2^-$ state. 
The $D$-wave nature of the $3/2^-$ coupling, with its associated Legendre polynomials $P_2^0(\cos\theta)$ and $P_2^1(\cos\theta)$ for the non-spin-flip and spin-flip amplitudes respectively, yields an angular distribution of the form $a + b\cos^2\theta+c\cos^4\theta$ for the pure resonance contribution. 
Combined with other mechanisms, it produces the shapes observed in the data. 
Some interesting differences between the two solutions appears in the $K^0\Xi^0$ channel. 
As shown in Fig.~\ref{fig:DSC_neutral}, at most energy points, the curve shape of Solution‑II is smoother than that of Solution‑I, because of distinct interference patterns between the $3/2^-$ hyperon and other particles. 
However, at $\sqrt{s}=2.28$~GeV the two solutions predict noticeably different angular distributions: Solution-I exhibits a flatter shape, whereas Solution-II exhibits a more structured angular dependence. 
This difference reflects the distinct relative phases between the $3/2^-$ state and $\Sigma(2250)$  in the two solutions (see Tables~\ref{tab:1} and~\ref{tab:2}), which alter the interference pattern among the $s$-channel resonances.
The existing experimental data at this energy are too sparse to distinguish between the two predictions, but future high-statistics measurements could exploit this difference.

In summary, both the total and differential cross sections are consistent with the $J^P=3/2^-$ assignment of the newly introduced state.
However, comparing the two solutions shows that they predict nearly identical cross sections for most observables, with the $K^0\Xi^0$ differential cross section at 2.28~GeV being a notable exception.
The slight difference in $\chi^2/\mathrm{d.o.f.}$ between Solution-I ($3.05$) and Solution-II ($3.20$) is not statistically significant given the uncertainties in the data. 
This degeneracy arises because the cross sections are not sufficiently sensitive to the detailed isospin structure of the $3/2^-$ state. 
To break this degeneracy, one must examine observables that are more responsive to the isospin-dependent interference patterns, such as $\Xi$ polarization and the $K^-n$ channel, as discussed in the following subsections.

\subsection{$\Xi$ polarization}

\begin{figure}[tb]
    \centering
    \begin{subfigure}
        \centering
        \includegraphics[width=0.48\linewidth]{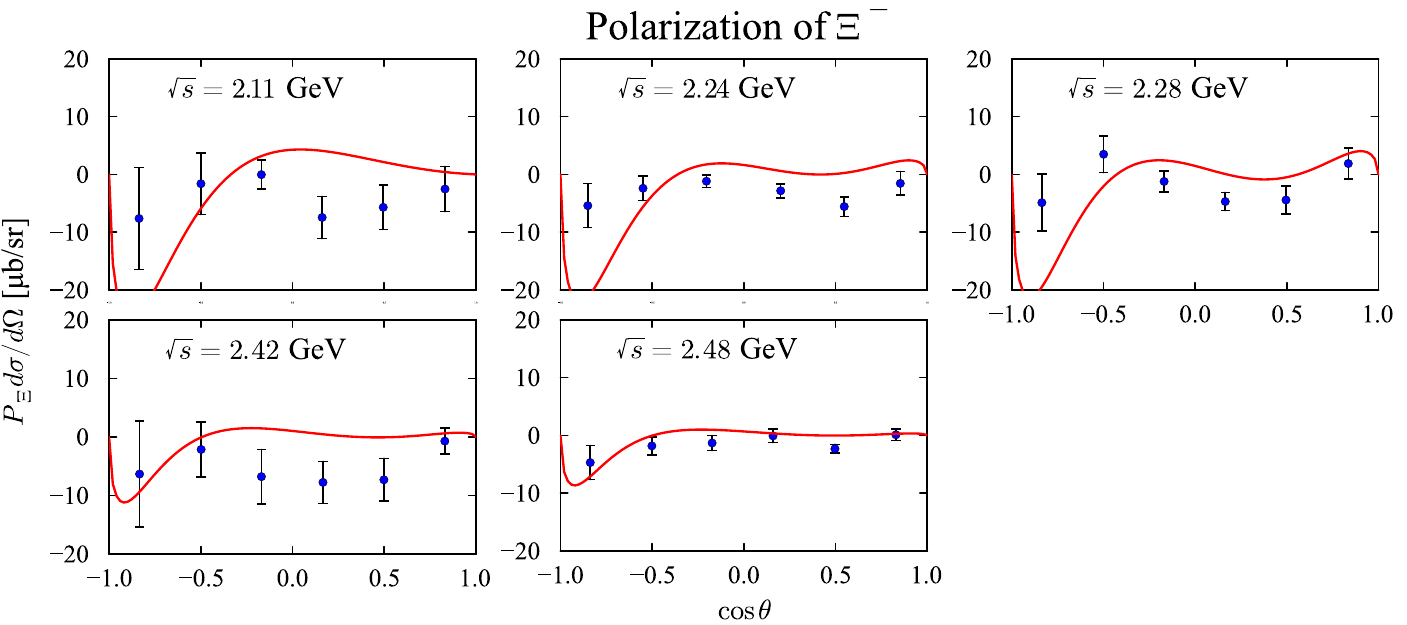}
    \end{subfigure}
    \hfill
    \begin{subfigure}
        \centering
        \includegraphics[width=0.48\linewidth]{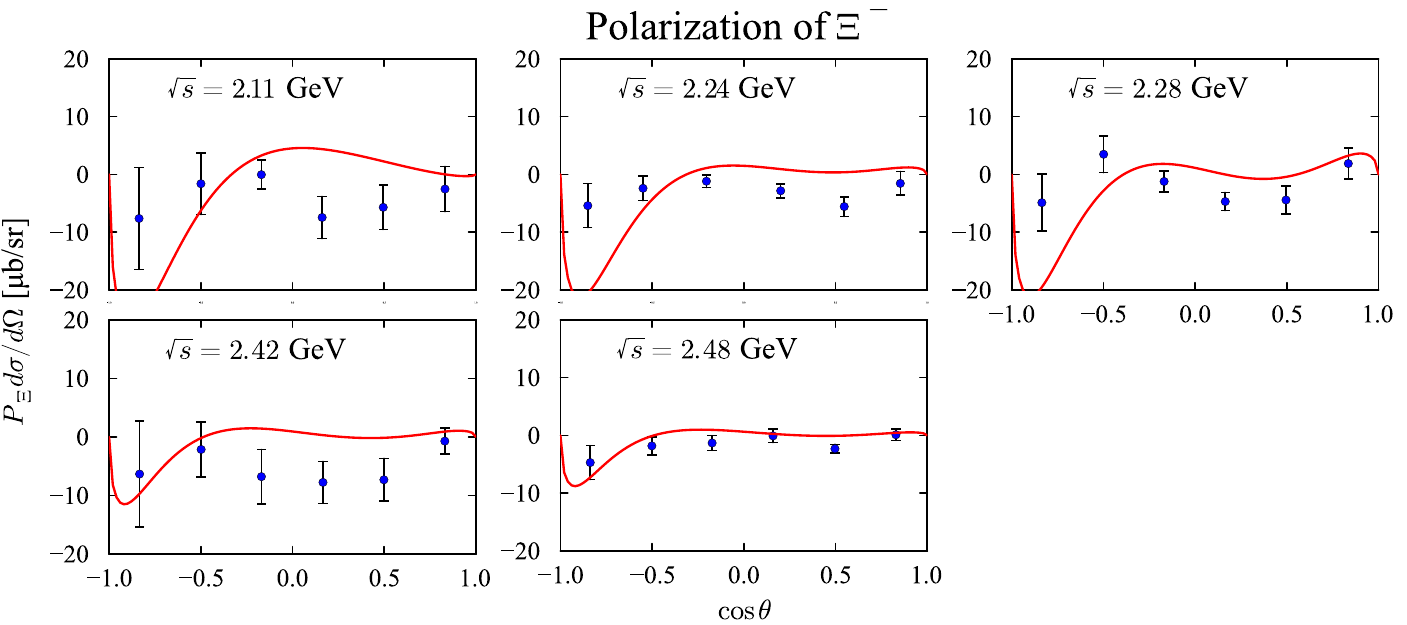}
    \end{subfigure}
    \vspace{1em}
    \begin{subfigure}
        \centering
        \includegraphics[width=0.48\linewidth]{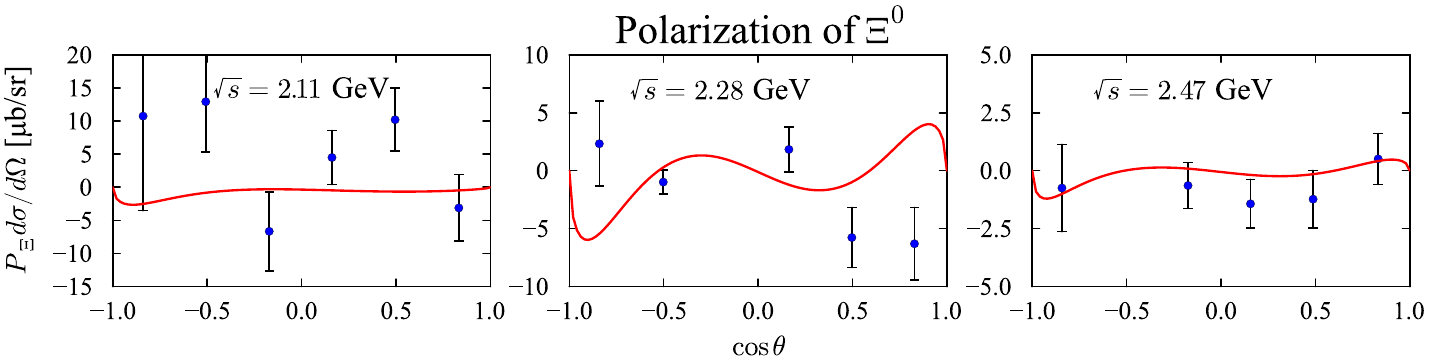}
    \end{subfigure}
    \hfill
    \begin{subfigure}
        \centering
        \includegraphics[width=0.48\linewidth]{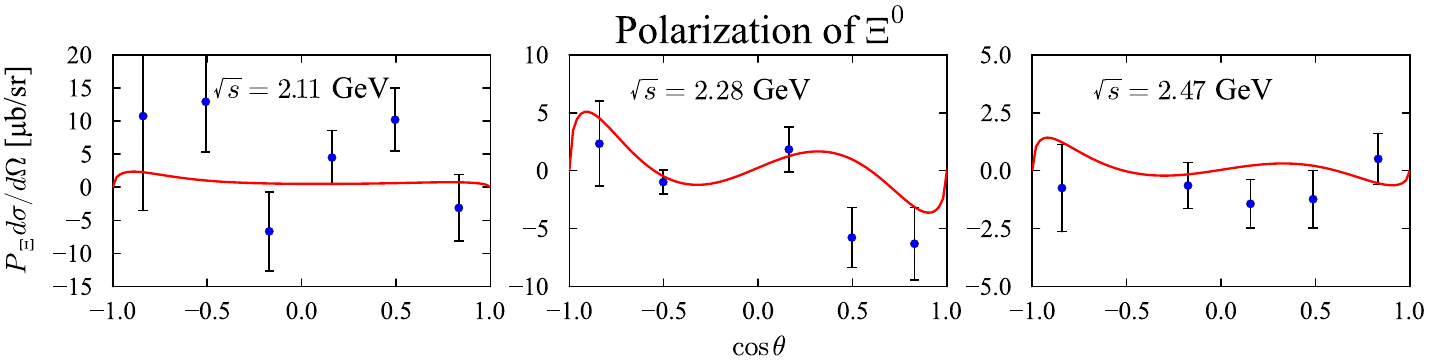}
    \end{subfigure}
    \caption{Polarizations of $\Xi^-$ (top row) and $\Xi^0$ (bottom row) in the $K^+\Xi^-$ and $K^0\Xi^0$ channels, respectively, for unpolarized initial states. Solution-I results are shown in the left column and Solution-II in the right column. The scattering angle $\theta$ is defined as the angle between the direction of the incoming $K^-$ and the outgoing $K$ in the c.m.\ frame. The experimental data are taken from Refs.~\cite{Trippe:1967wat,Dauber:1969hg}.}
    \label{fig:Polar}
\end{figure}

The polarization of the final-state $\Xi$ provides a more sensitive probe of the reaction mechanism, as it arises from the interference between different spin amplitudes and is therefore particularly responsive to the relative phases and isospin structures of the contributing resonances. 
We evaluate the polarization of $\Xi^-$ and $\Xi^0$ as an independent test of the two solutions, as shown in Fig.~\ref{fig:Polar}. 

In the $K^+\Xi^-$ channel, both solutions predict rather similar polarization patterns and reproduce the overall trend of experimental data. 
The distinction between the two solutions becomes more apparent in the $K^0\Xi^0$ polarization at $\sqrt{s}=2.28$~GeV, where Solution-II predicts a polarization that is closer to the data points in angular dependence, while Solution-I shows larger deviations. 
Overall, Solution-II with the $\Lambda^*(3/2^-)$ assignment provides a somewhat better description of the available polarization data.

The difference in the polarization of two solutions stems from the distinct isospin structures of the two scenarios and the associated phase factors. 
Since the polarization is proportional to the imaginary part of the interference between different spin-flip and non-spin-flip amplitudes, it is sensitive to relative phases among the contributing resonances, which differ between the two solutions.

Given that the polarization data, despite their rather large uncertainties, show a modest preference for Solution-II, this observable provides a useful constraint for distinguishing between the two scenarios. 
However, the limited precision and sparse energy coverage of the existing polarization measurements preclude a definitive conclusion. 
High-statistics polarization measurements at future facilities such as J-PARC~\cite{Aoki:2021cqa} and HIAF~\cite{Chen:2025ppt} would be highly valuable for confirming or refuting this indication.

\subsection{Predictions for $K^-n\to K^0\Xi^-$}

\begin{figure}[tb]
    \centering
    \begin{subfigure}
        \centering
        \includegraphics[width=0.48\linewidth]{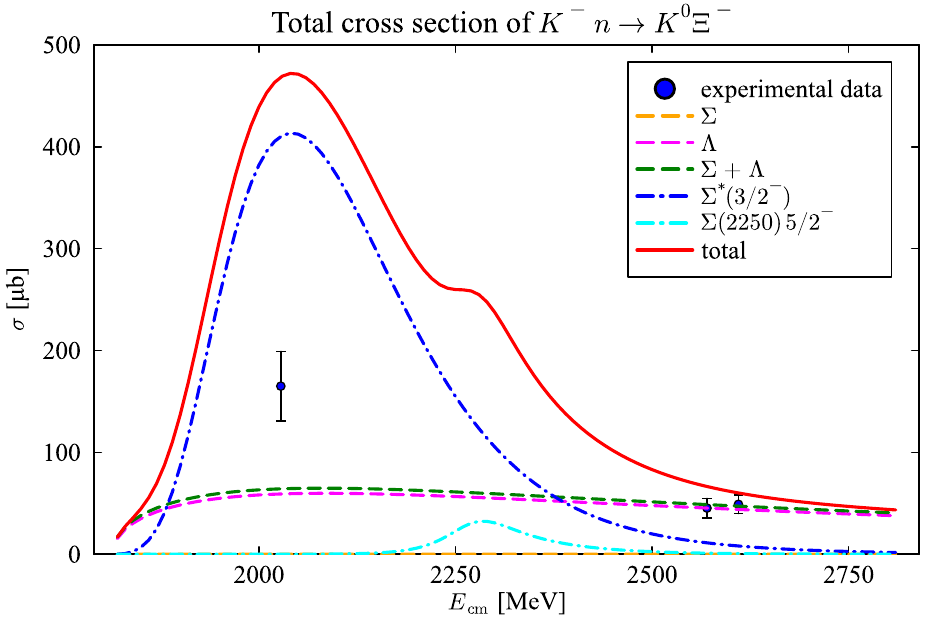}
    \end{subfigure}
    \hfill
    \begin{subfigure}
        \centering
        \includegraphics[width=0.48\linewidth]{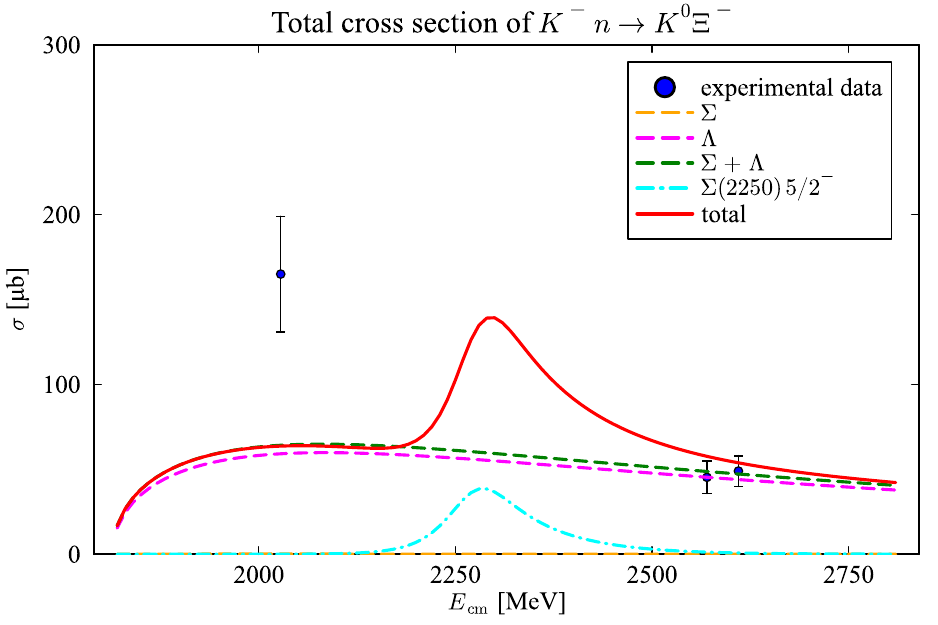}
    \end{subfigure}
    \vspace{1em}
    \begin{subfigure}
        \centering
        \includegraphics[width=0.48\linewidth]{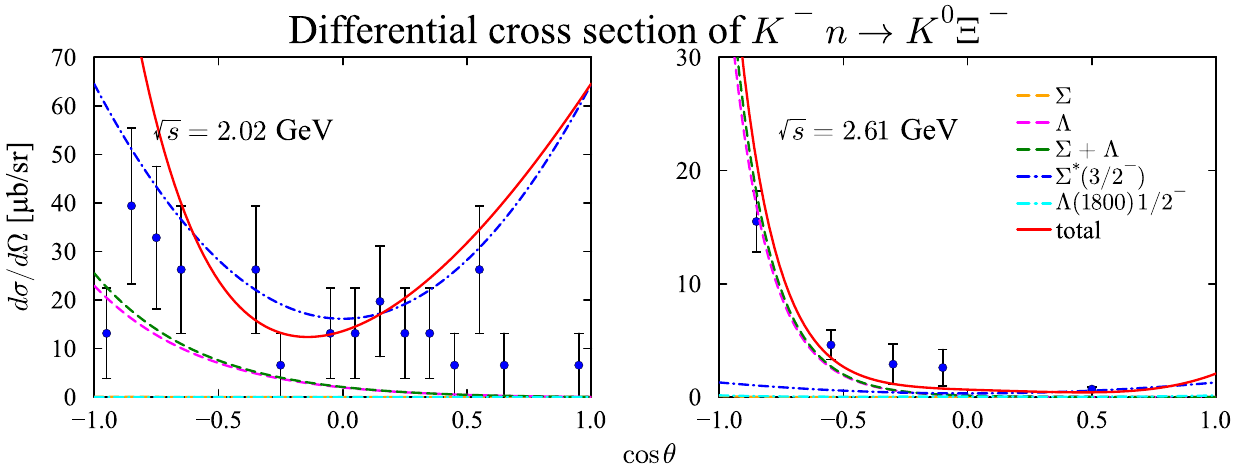}
    \end{subfigure}
    \hfill
    \begin{subfigure}
        \centering
        \includegraphics[width=0.48\linewidth]{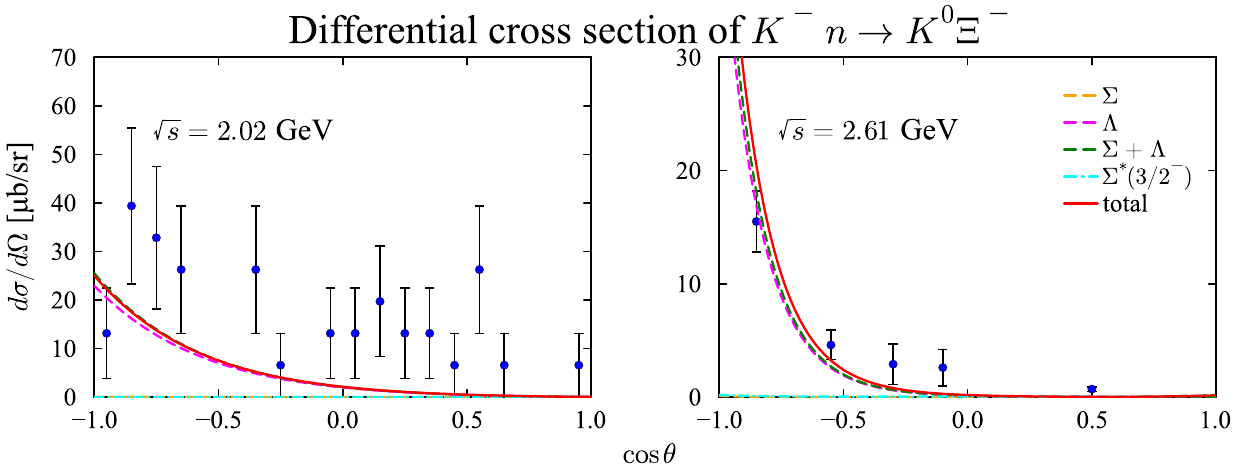}
    \end{subfigure}
    \caption{Total (top row) and differential (bottom row) cross sections of $K^-n\to K^0\Xi^-$ predicted by Solution-I (left column) and Solution-II (right column). The scattering angle $\theta$ is defined as the angle between the direction of the incoming $K^-$ and the outgoing $K^0$ in the c.m.\ frame. The experimental data of total cross section are taken from Refs.~\cite{Berge:1966zz,SABRE:1971pzp,Briefel:1975tm}, while the data of differential cross section are taken from Refs.~\cite{Berge:1966zz,SABRE:1971pzp}.}
    \label{fig:K_n}
\end{figure}

As another independent test of the two scenarios, we present predictions for the $K^-n\to K^0\Xi^-$ reaction in Fig.~\ref{fig:K_n}. 
This channel has a different isospin decomposition than the $K^-p$ initiated reactions, making it particularly sensitive to the isospin of the intermediate resonances. 
The amplitudes can be obtained from the isospin relation $\mathcal{M}_{K^-n\to K^0\Xi^-} = \mathcal{M}_{K^-p\to K^+\Xi^-} - \mathcal{M}_{K^-p\to K^0\Xi^0}$, which leads to distinct interference patterns for the two solutions.

The total cross section predictions from the two solutions show noticeable differences, especially in the peak magnitude around $\sqrt{s}\approx 2.0$~GeV. 
In Solution-I, where the $3/2^-$ state is a $\Sigma^*$ with isospin $I=1$, the resonance contributes to the neutron channel with an isospin factor ($-2$) that differs from that in the proton channels ($-1$ or 1). 
However, in Solution-II, the state $\Lambda^*(3/2^-)$ with $I=0$ does not contribute to the neutron channel. 
Consequently, the predicted cross section in Solution-I is generally larger than that in Solution-II, with the difference being most pronounced near the resonance peak. 

The differential cross sections also exhibit distinct angular dependence in the two scenarios. 
Solution-II predicts a shape that is in better agreement with the angular distribution of the experimental data, although its overall magnitude is slightly lower than the data. 
At $\sqrt{s}=2.02$~GeV, Solution-I, while closer to the data at intermediate angles, exhibits a more pronounced forward and backward enhancement that deviates from the measured angular trend.
This occurs near the resonance peak, where the $s$-channel contribution is dominant, making it a useful discriminant between the two scenarios.

We emphasize that the $K^-n\to K^0\Xi^-$ channel provides an excellent discriminator between the two solutions. 
As can be seen from the isospin decomposition, $\Lambda$-type $s$-channel resonances cancel out in this channel, while only $\Sigma$-type $s$-channel exchanges contribute. 
Therefore, if the $J^P=3/2^-$ state is a $\Sigma^*$, it produces a significant signal in this channel, whereas if it is a $\Lambda^*$, it does not contribute to the $s$-channel. 
This makes the $K^-n\to K^0\Xi^-$ reaction a particularly clean probe of the isospin of the $3/2^-$ state. 
Future measurements of this reaction with improved statistics, as planned at J-PARC~\cite{Aoki:2021cqa} and HIAF~\cite{Chen:2025ppt}, will therefore provide a crucial test to distinguish between the $\Sigma^*(3/2^-)$ and $\Lambda^*(3/2^-)$ hypotheses. 
Combined with precision polarization measurements in the $K^-p$ channels, such data should be able to conclusively determine the isospin of the $J^P=3/2^-$ state.

\subsection{On the nature of the $J^P=3/2^-$ state}

Having shown that both the $\Sigma^*(3/2^-)$ and $\Lambda^*(3/2^-)$ assignments can describe the existing data, with a slight preference for the $\Lambda^*(3/2^-)$ one from polarization observables and $K^-n\to K^0\Xi^-$ data, we briefly discuss the possible nature of these two states. 
In the relativistic quark model predictions of Ref.~\cite{Capstick:1986ter}, neither a $\Sigma^*(3/2^-)$ nor a $\Lambda^*(3/2^-)$ state with a mass near 1900~MeV is present in the spectrum. 
The closest $3/2^-$ states in the quark model lie at significantly higher or lower masses. 
Since such a $3/2^-$ states could not be explained as a conventional three-quark state, it might serves as a candidate for a $s\bar{s}qqs$ pentaquark hyperon. Whether such pentaquark components are compact or molecular in nature requires further investigation.

Considering the substantial uncertainties from our fitting procedure (the mass is $1890 \pm 220$~MeV for solution-I, $1890 \pm 180$~MeV for solution-II), the mass of this state could be close to the threshold of $K\Xi(1530)$, where $\Xi(1530)$ is the well-established $\Xi^*$ with $J^P=3/2^+$. 
This proximity suggests the possibility of an $S$-wave $K\Xi(1530)$ hadronic molecule interpretation. 
In the molecular picture, the $\Sigma^*(3/2^-)$ and $\Lambda^*(3/2^-)$ would correspond to the $I=1$ and $I=0$ combinations of the $K\Xi(1530)$ system, respectively, and both could in principle exist as quasi-bound states or virtual states near threshold. Note that comparing with the case of $\bar D\Xi_c^*$ system~\cite{Dong:2021bvy,Dong:2021juy} as its $c\bar{c}qqs$ pentaquark partner, the $K\Xi(1530)$ system has stronger attractive force due to the additional $\phi$ meson exchange.  
If this interpretation is correct, the partner state (e.g., the $\Sigma^*(3/2^-)$, if the $\Lambda^*(3/2^-)$ is confirmed, or vice versa) should also exist and could be searched for in dedicated experiments. 

In this energy region, the $K\Xi(1530)$ channel can couple with $\Sigma(1385)\eta$ and $\Lambda\phi$ to form coupled-channel systems with $I=0$ and $I=1$, $J^P=3/2^-$ hyperon states. Such two-body coupled-channel systems may dynamically generate resonance poles, analogous to the $\Lambda(1405)$ and its two-pole structure arising from $\Sigma\pi$ and $\bar{K}N$ coupling. Further theoretical and experimental studies based on the present work will deepen our understanding of the $s\bar{s}qqs$ components in hyperon excitations.

\section{Summary}\label{sec4}

We have analyzed the $K^-p \to K\Xi$ reaction within an effective Lagrangian approach by simultaneously analyzing the $K^+\Xi^-$ and $K^0\Xi^0$ channels. Our model includes the ground-state $\Sigma$ and $\Lambda$ hyperons, $\Lambda(1800)$ and $\Sigma(2250)$ resonances, and a possible $J^P = 3/2^-$ hyperon with strangeness $S=-1$ with mass around 1.9$\,$GeV and width about 200$\,$MeV. This state can be either a $\Sigma^*(3/2^-)$ (Solution-I) or a $\Lambda^*(3/2^-)$ (Solution-II). Both scenarios describe the data
comparably well. The newly introduced $3/2^-$ state significantly improves the description of the
experimental data compared to previous analyses. 

Future experiments at J-PARC~\cite{Aoki:2021cqa}, HIAF~\cite{Chen:2025ppt} and JLab~\cite{Dobbs:2022agy} will provide excellent opportunities to test and distinguish the two solutions. Specifically, the $\Sigma^*(3/2^-)$ can be probed via the $K^-n \to K^0\Xi^-$ and $K_Lp\to K^+\Xi^0$ reaction, while the $\Lambda^*(3/2^-)$ can be examined through the $K^-p \to \pi^0\Sigma^0$ channels. Together with the predictions for $\Xi$ polarization presented in this work, such measurements will provide a stringent test of the nature of the $J^P = 3/2^-$ hyperon and deepen our understanding of the baryon excitation spectrum in the 2~GeV region.

The interpretation of this $J^P = 3/2^-$ hyperon with $S=-1$ has been discussed in Sec.~\ref{sec3}. The absence of such a state in the relativistic quark model predictions~\cite{Capstick:1986ter} suggests that it could be an $s\bar{s}qqs$ pentaquark hyperon, while its proximity to the $K\Xi(1530)$ threshold hints that it might be interpreted as an $S$-wave $K\Xi(1530)$ hadronic molecule.
If confirmed, this would enrich our understanding of the baryon spectrum beyond the conventional three-quark picture.

\begin{acknowledgments}
We are grateful to Feng-Kun Guo, Shu-Ming Wu and Hao-Jie Jing for useful discussions.
This work is supported by the National Natural Science
Foundation of China (Grants No.12547111, No.12221005), and the Chinese Academy of Sciences under Grant No. YSBR-101, and the National Key Research and Development Program of China under Contract No. 2025YFA1613900.
\end{acknowledgments}

\appendix
\section{Conventions on isospin multiplets}\label{app:1}
The notations of isospin doublets are
\begin{equation}
    N=\begin{pmatrix}
        p\\
        n
    \end{pmatrix},\quad
    \Xi=\begin{pmatrix}
        \Xi^0\\
        -\Xi^-
    \end{pmatrix},\quad
    K=\begin{pmatrix}
        K^+\\
        K^0
    \end{pmatrix},\quad
    K_c=\begin{pmatrix}
        \bar K^0\\
        -K^-
    \end{pmatrix},
\end{equation}
and the inner product of Pauli matrix $\bm\tau$ and isospin triplet $\bm\Sigma$ is
\begin{equation}
    \bm{\tau}\cdot\bm{\Sigma}=
    \begin{pmatrix}
        \Sigma^0 & \sqrt{2}\Sigma^+\\
        \sqrt{2}\Sigma^- & -\Sigma^0
    \end{pmatrix}.
\end{equation}

\section{Amplitudes}\label{app:2}
The amplitudes involved in this work are listed as follow:
\begin{equation}
    \mathcal M^s_\Sigma=g_{\Sigma\Xi K}g_{\Sigma N\bar K}F_{\Sigma}^2(s)\bar u_\Xi(k_2,r_2)\gamma^5\left(\lambda+\frac{1-\lambda}{m_\Xi+m_\Sigma}\slashed k_1\right)\frac{\slashed p_1+\slashed p_2+m_\Sigma}{s-m_\Sigma^2}\left(\lambda+\frac{1-\lambda}{m_N+m_\Sigma}\slashed p_1\right)\gamma^5 u_N(p_2,s_2),
\end{equation}
\begin{equation}
    \mathcal M^u_\Sigma=g_{\Sigma\Xi K}g_{\Sigma N\bar K}F_{\Sigma}^2(u)\bar u_\Xi(k_2,r_2)\gamma^5\left(\lambda-\frac{1-\lambda}{m_\Xi+m_\Sigma}\slashed p_1\right)\frac{\slashed p_2-\slashed k_1+m_\Sigma}{u-m_\Sigma^2}\left(\lambda-\frac{1-\lambda}{m_N+m_\Sigma}\slashed k_1\right)\gamma^5 u_N(p_2,s_2),
\end{equation}
\begin{equation}
    \mathcal M^s_\Lambda=g_{\Lambda\Xi K}g_{\Lambda N\bar K}F_{\Lambda}^2(s)\bar u_\Xi(k_2,r_2)\gamma^5\left(\lambda+\frac{1-\lambda}{m_\Xi+m_\Lambda}\slashed k_1\right)\frac{\slashed p_1+\slashed p_2+m_\Lambda}{s-m_\Lambda^2}\left(\lambda+\frac{1-\lambda}{m_N+m_\Lambda}\slashed p_1\right)\gamma^5 u_N(p_2,s_2),
\end{equation}
\begin{equation}
    \mathcal M^u_\Lambda=g_{\Lambda\Xi K}g_{\Lambda N\bar K}F_{\Lambda}^2(u)\bar u_\Xi(k_2,r_2)\gamma^5\left(\lambda-\frac{1-\lambda}{m_\Xi+m_\Lambda}\slashed p_1\right)\frac{\slashed p_2-\slashed k_1+m_\Lambda}{u-m_\Lambda^2}\left(\lambda-\frac{1-\lambda}{m_N+m_\Lambda}\slashed k_1\right)\gamma^5 u_N(p_2,s_2),
\end{equation}

\begin{equation}
    \mathcal M^s_{\Lambda(1/2^-)}=-g_{\Lambda(1/2^-)\Xi K}g_{\Lambda(1/2^-) N\bar K}F_{\Lambda(1/2^-)}^2(s)\bar u_\Xi(k_2,r_2)\frac{\slashed p_1+\slashed p_2+m_{\Lambda(1/2^-)}}{s-m_{\Lambda(1/2^-)}^2+im_{\Lambda(1/2^-)}\Gamma_{\Lambda(1/2^-)}} u_N(p_2,s_2),
\end{equation}
\begin{equation}
    \mathcal M^s_{\Lambda(3/2^-)}=\frac{g_{\Lambda(3/2^-)\Xi K}g_{\Lambda(3/2^-) N\bar K}}{m_K^2}F_{\Lambda(3/2^-)}^2(s)\bar u_\Xi(k_2,r_2)\gamma^5\frac{(\slashed p_1+\slashed p_2+\sqrt{s})\Delta_{\mu\nu}k_1^\mu p_1^\nu}{s-m_{\Lambda(3/2^-)}^2+im_{\Lambda(3/2^-)}\Gamma_{\Lambda(3/2^-)}}\gamma^5 u_N(p_2,s_2),
\end{equation}
\begin{equation}
    \mathcal M^s_{\Sigma(3/2^-)}=\frac{g_{\Sigma(3/2^-)\Xi K}g_{\Sigma(3/2^-)N\bar K}}{m_K^2} F_{\Sigma(3/2^-)}^2(s)\bar u_\Xi(k_2,r_2)\gamma^5\frac{(\slashed p_1+\slashed p_2+\sqrt{s})\Delta_{\mu\nu}k_1^\mu p_1^\nu}{s-m_{\Sigma(3/2^-)}^2+im_{\Sigma(3/2^-)}\Gamma_{\Sigma(3/2^-)}}\gamma^5 u_N(p_2,s_2),
\end{equation}
\begin{equation}
    \mathcal M^s_{\Sigma(5/2^-)}=-\frac{g_{\Sigma(5/2^-)\Xi K}g_{\Sigma(5/2^-) N\bar K}}{m_K^4}F_{\Sigma(5/2^-)}^2(s)\bar u_\Xi(k_2,r_2)\frac{(\slashed p_1+\slashed p_2+\sqrt{s})\Delta_{\mu\nu,\alpha\beta}k_1^\mu k_1^\nu p_1^\alpha p_1^\beta}{s-m_{\Sigma(5/2^-)}^2+im_{\Sigma(5/2^-)}\Gamma_{\Sigma(5/2^-)}} u_N(p_2,s_2),
\end{equation}
where $s=(p_1+p_2)^2$, $u=(p_2-k_1)^2$, and $s_2$, $r_2$ are the spin indices of the initial $N$ and final $\Xi$, respectively.
\bibliography{refs}
\end{document}